\documentclass[aps,pre,twocolumn,showpacs,amssymb,floatfix,superscriptaddress,letter]{revtex4-1}

\usepackage{amssymb}
\usepackage{latexsym}
\usepackage{amsfonts}
\usepackage{epsfig}
\usepackage{amsmath}
\usepackage{float}  
\usepackage{verbatim}
\usepackage{graphicx}
\usepackage{color}

\newcommand{\hl}[1]{{#1}} 

\begin{document}

\title{Assortativity Decreases the Robustness \\
 of Interdependent Networks}

\author{Di Zhou}
\affiliation{Center for Polymer Studies and Department of Physics,
Boston University, Boston, MA 02215 USA}

\author{Gregorio D'Agostino}
\affiliation{ENEA - CR ``Casaccia'' - via Anguillarese 301 I-00123 Roma, Italy}

\author{Antonio Scala}
\affiliation{ISC-CNR Dipartimento di Fisica, Sapienza Universit\`a di Roma
Piazzale Moro 5, 00185 Roma, Italy}
\affiliation{London Institute of Mathematical Sciences, 22 South Audley St
Mayfair London W1K 2NY, UK}

\author{H. Eugene Stanley}
\affiliation{Center for Polymer Studies and Department of Physics,
Boston University, Boston, MA 02215 USA}

\date{today}

\begin{abstract}
It was recently recognized that
interdependencies among different networks can play a crucial role in
triggering cascading failures and hence system-wide disasters. A recent
model shows how pairs of interdependent networks can exhibit an abrupt
percolation transition as failures accumulate.  We report on the effects
of topology on failure propagation for a model system consisting of two
interdependent networks.  We find that the internal node correlations in
each of the two interdependent networks significantly changes the
critical density of failures that triggers the total disruption of the
two-network system. \hl{ Specifically, we find that the assortativity
(i.e. the likelihood of nodes with similar degree to be connected)}
within a single network decreases the robustness of the entire 
system.
The results of this study on the influence of assortativity may provide
insights into ways of improving the robustness of network architecture,
and thus enhance\hl{s} the level of protection of critical
infrastructures.

\end{abstract}


\maketitle

\section{Introduction}

The quality of life in modern society strongly depends on the effective
delivery of basic services as water, electricity, and
communications; the infrastructures providing these basic services
are therefore called critical infrastructures. Indeed, maintaining every critical
infrastructure (CI) is a growing challenge for modern society. Of great
interest is the interdependence of CIs. One clear example of this
interdependence is the ``binomial'' system in which electrical power
networks depend on telecommunication networks and {\em vice-versa}.
Understanding cascading failure in interdependent networks is a problem
currently receiving much attention.  For example, failure propagation is
a common phenomenon that can lead to such catastrophic effects as the
remarkable September 2003 total blackout across Italy
\cite{BerizziIEEE2004}.

One approach to understanding failure propagation is to develop a
simulation of an entire system that takes into account all of the details
associated with the system. Although some remarkable results have been
achieved for selected regions when all the information is available
\cite{DIESIS}, these simulations are not useful in understanding the
mechanisms that induce cascading effects. Because of the huge amount of
data involved, such an approach requires heroic efforts, even when
considering a simple system \cite{IRRIIS}. Frequently privacy
constraints or difficulties in accessing or probing the system of
interest become factors, and often an adequate amount of information is
not available.  Thus other approaches are needed to understand the
fundamental issues underlying cascading effects.

Here we will use the complex network paradigm to acquire some understanding
of possible CI vulnerabilities to help focus more detailed analyses
\cite{KroegerZioBOOK}.  For example, the complex network paradigm can
address the problem of interdependencies among CIs.  It was recently
demonstrated \cite{BrummitPNAS2012} \hl{ that when the degree of
interdependence between networks is increased, 
the robustness of the system to cascade failures decreases 
} \cite{BuldyrevNAT2010} \hl{and \textit{vice-versa}}.

 Beside the natural applications to the protection of CIs, the analysis 
of the properties of inter-dependent networks is a subject of growing interest
in many scientific fields, ranging from public cooperation \cite{WangEPL2012}, 
to epidemic spreading \cite{JolandAPS2012,DickisonAPS2012,SaumellPRE2012,SonEPL2012}, 
to human physiology \cite{PlamenNAT2012}.

Although previous research has indicated how to mitigate systemic risk
by tuning network interdependencies \cite{NewmanHaway2005} , the role of topology of each
component network has yet to be investigated. The interdependency level
among a set of networks can be fixed---because of economic or
technological constraints---and thus may be untunable.

If we know the average degree of a network, degree correlation becomes
the simplest parameter for classifying the internal network topology.
Using the interacting failure model (IFM) for a two-layered network
system 2LNS \cite{BuldyrevNAT2010}, we find that assortativity decreases
the robustness to random failure.  In particular, by considering both
Erd\H{o}s-R\'{e}nyi (ER) and scale-free (SF) networks, we find that assortativity
causes a sharp increase in the fragility of coupled networks with a node
distribution described by a power law.
 
In Sec.~\ref{sec:methods} we review the concept of assortativity,
describe the methods employed to generate sample networks with different
assortativity coefficients, and review the two-layer network model of
cascading failures.  In Sec.~\ref{sec:results} we present the results of
our simulations, which we then discuss in Sec.~\ref{sec:conclusions}.

\section{Methods}\label{sec:methods}

A fundamental quantity characterizing the structure and driving the
behavior of a large network is the probability distribution function
$P(k)$ of node degree $k$ \cite{CaldarelliBOOK2007,BarratBOOK2008}.  It
has been shown that both humanly-constructed and natural networks are
often characterized by a $P(k)$ with heavy tails leading to unforeseen
effects, e.g., the disappearance of epidemic thresholds
\cite{PSatorrasPRL01}.  Other quantities, e.g., the local density of
triangles, the modular structure, communities, and motifs can be used to
further characterize network structure \cite{CaldarelliBOOK2007}.

Assortativity is the tendency of entities to seek out and group with
those other entities that exhibit similar characteristics.  In networks,
assortativity is the tendency of neighbor nodes to have similar degrees
and thus to be measurable using link-averaged degree pair correlation
\cite{NewmanPRL2002}. Using a physical approach, we improve the
one-point average-degree characterization of a network by considering
assortativity, a two-point correlation quantity.

The assortativity coefficient $r$ is defined in terms of the 
correlation between the degrees of neighbouring vertices
\begin{equation}
\label{eq:r}
 r \equiv \frac{\langle jk\rangle_e -  \left[ \langle \left( j+k \right)/2 \rangle_e \right]^2}
 {\langle (j^2+k^2)/2 \rangle_e - \left[ \langle (j+k)/2 \rangle_e \right]^2},
\end{equation}
where the averages $\langle\cdots\rangle_e$ are evaluated over all
edges $e$ and $j,k$ are the degrees of the adjacent vertices associated
with edge $e$. 
\hl{High values of the assortativity ($r\sim 1$) imply that neighbouring nodes have similar degrees, while low values ($r\sim -1$)
imply that high-degree nodes tend to be connected to low-degree ones; 
random pairing corresponds to $r\sim 0$.}

\hl{An alternative form for this expression has been given by Newman}
\cite{NewmanPRL2002} 
\begin{equation}
\label{eq:r_newman}
 r = \frac{ \sum_{jk} jk \left( e_{jk} - q_j q_k \right) }
 { \sum_{k}  \left[ k^2 q_k - \left( k q_k \right)^2 \right] }  
\end{equation}
\hl{in terms of the normalized degree distribution 
$q_k = \left(k+1\right)P(k+1)/\sum_k k P(k) $ 
and the joint probability distribution $e_{ij}$ 
(i.e. the two point function) 
of the residual degrees
at the either ends of a randomly chosen edge.}

For a single stand-alone network, increasing the assortativity makes it
more robust against node removal \cite{NewmanPRE2003,VazquezPRE2003}
and, in general, stronger with respect to diffusion-driven dynamical
processes \cite{DAgostinoEPL2012}.  On the other hand, assortativity
makes networks more unstable \cite{BredeARX2005} as measured by the May
criterion \cite{MayBOOK1973} and, in general, less controllable
\cite{MieghemBOOK2010,DAgostinoEPL2012}.

\subsection{Varying the assortativity}

We next demonstrate how assortativity affects cascading fault propagation
in interdependent networks by tuning $r$ while keeping the degree
distribution fixed. 
To produce networks of varying assortativity, several methods (mostly based on link swapping \cite{Maslov2002} 
have been employed, like accepting assortative moves with a given probability $p$\cite{XulviBrunetPRE2004}. 
We believe that the most flexible way of sampling of 
the space of possible networks is to introduce a simple 
Hamiltonian\cite{NohPRE2007,DAgostinoEPL2012}
as it allows to apply all the standard tools of statistical mechanics.

Following Ref.~\cite{DAgostinoEPL2012}, we first define {\bf A}, the
adjacency matrix associated with the network (an $N\times N$ matrix
exhibiting a unitary value $A_{ij}=1$ when node $i$ is linked to node
$j$ and vanishing elsewhere). Next, we define a statistical ensemble in
which the probability measure $\mu(G)\propto \exp{[-J\,H(G)]}$ of a
given graph $G$ is induced by the Hamiltonian
\begin{equation}
{\cal H}(G)\equiv\sum_{ij}k_{i}A_{ij}k_{j}.
\end{equation}
Here $k_i$ is the degree of node $i$, and the ``coupling constant'' $J$
may assume both positive and negative values. 
Such Hamiltonian is a simple quadratic form in the node degrees resembling
strictly the form applied to model classical spin systems.

Note that for a given $P(k)$ the terms $\langle (j+k)/2 \rangle_e$ and
$\langle (j^2+k^2)/2 \rangle_e$ in Eq.~(\ref{eq:r}) do not depend on how
nodes are linked, but that the term $\langle jk\rangle_e$ is directly
related to the Hamiltonian as ${\cal H} \propto \langle k_i k_j
\rangle$. Thus large values of $J$ will favour graphs with large values
of $r$ and \textit{vice-versa}.

\begin{figure}
\begin{center}
\includegraphics[width=0.95 \columnwidth]{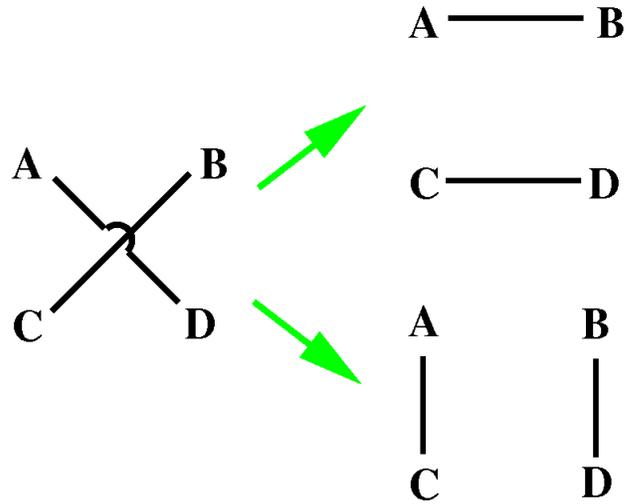}
\end{center}
\caption{ \hl{ (Color online) The link-swap procedure consists 
in deleting two edges $AB$ and $CD$ and adding either the 
edges $AC$,$BD$ or the edges $AD$,$BD$ respecting constrains like 
the absence of multiple links. Such a procedure leaves always the 
number of links attached to each node unchanged and therefore 
it leaves the degrees of the nodes unchanged.}
}
\label{Fig1}
\end{figure}

Such a Hamiltonian has been recently explored by Yook and Park\cite{YookEPL2011}
who have found that the network configuration sampled at equilibrium satisfy
the power law distribution $P(k)\sim k^{-3/2}$.
We instead explore the configuration space defined by link swapping: 
in this way not only the initial $P(k)$ but also each node degree is kept constant \hl{(see fig.}\ref{Fig1}\hl{)}. 
In order to
sample configurations according to $\mu(G)$, we set the link swapping
probability to be $e^{-J,\Delta {\cal H}}$.  Although link-swap moves can
be assortative, disassortative, or neutral \cite{Zhou2007}, in our case
assortative/disassortative configurations will be preferentially sampled
according to the sign of $J$ with a monotonically increasing sampling of
the assortativity $r$ with respect to the parameter $J$
\cite{DAgostinoEPL2012}.

In order to examine the effects of assortativity for both SF and ER
networks, we use both the Barabasi-Albert (BA) \cite{BAreview} and the
ER \cite{BollobasBOOK1998} model networks as starting configurations for
link-swapping Monte Carlo (MC) dynamics. We find that in ER networks MC
equilibrium is reached with a number of steps per node apparently
independent of the number of nodes \cite{NohPRE2007}, but that in SF
networks the situation is more complicated. We find that in BA networks
the range of assortativities reached in a given number of steps shrinks
as the system size increases \cite{MenchePRE2010}.  Thus similar
allocated simulation times allow us to explore a smaller assortativity
range for BA networks than for ER networks.

\subsection{Interdependent failures model}

To model interdependent networks with assortativity we use the
interdependent failure model (IFM) of Ref.~\cite{BuldyrevNAT2010}. In
the original IFM there are two spatial networks (ICT and power
distribution) and only the geographically nearest nodes interact. We
consider two networks $A$ and $B$ that have the same number of nodes $N$
and that share the same topology, and thus the same assortativity (or
disassortativity).  We consider a case in which a unique node $b_i$ in
network $B$ corresponds to a unique node $a_i$ in network $A$, i.e.,
$a_i$ and $b_i$ have a mutual dependence.  In the IFM, in order for node
$a_i$ to function properly, node $b_i$ must also function properly and
{\em vice-versa}. If $b_i$ becomes dysfunctional, $a_i$ will also become
dysfunctional.  This interdependence relation can be described as a
bidirectional link $a_i\leftrightarrow b_i$ between $a_i$ and
$b_i$. Thus each $a_i$ node in network $A$ has a corresponding
counterpart $b_i$ node in network $B$.

To model random attack or failure, we randomly remove a fraction $1-x$
of nodes from network $A$ ($x$ is the fraction of initially-surviving
nodes).  Because of the interdependence between the two networks, the
nodes in network $B$ that depend on the removed $A$-nodes are also
removed. When the nodes and links in network $B$ are removed, network
$B$ may break up into several connected components (``clusters'').  We
assume that only the nodes belonging to the largest cluster (the
so-called ``giant component'') continue to be functional
\cite{BuldyrevNAT2010}, and remove the nodes from $B$ that do not belong
to this giant component.  Because of mutual interdependency, removing
the $B$-nodes in network $B$ not in the giant component will cause the
removal of the corresponding $A$-nodes in network $A$. This iterative
process generates a cascade of failures that continues until node
elimination ceases. At that point, if the two networks still have giant
components, they will be the same size.  The algorithm used in this
procedure consists of the following steps:

\begin{list}{}{} 
\item [{~}]~
\item [{\texttt{\textit{\textcolor{blue}{s0:}}}}] Remove the 
fraction $1-x$ of initial failed nodes in layer $A$
\item [{\texttt{\textit{\textcolor{blue}{s1:}}}}] Identify the largest
component $S_{A}$ 
\item [{\texttt{\textit{\textcolor{blue}{s2:}}}}] Remove the $n_{A}$
nodes of $A$ not in $S_{A}$ 
\item [{\texttt{\textit{\textcolor{blue}{s3:}}}}] Remove the $n_{B}$
nodes of $B$ not linked to $S_{A}$
\item [{\texttt{\textit{\textcolor{blue}{s4:}}}}] Identify the largest
component $S_{B}$ 
\item [{\texttt{\textit{\textcolor{blue}{s5:}}}}] Remove the $n_{B}$
nodes of $B$ not in $S_{B}$ 
\item [{\texttt{\textit{\textcolor{blue}{s6:}}}}] Remove the $n_{A}$
nodes of $A$ not linked to $S_{B}$
\item [{\texttt{\textit{\textcolor{blue}{s7:}}}}] If $n_{A}>0$ then
repeat from \texttt{\textit{\textcolor{blue}{s1}}}
\item [{\texttt{\textit{\textcolor{blue}{s8:}}}}] Output the final survived
giant components $S_{A}$ and $S_{B}$
\label{alg:IFM}
\end{list}

Such algorithms can be re-phrased in terms of 
the fixpoint of an operator. In fact, let us define $Q_A$ as the operator that selects the largest component of network $A$ and let $P_{AB}$ 
the operator that select the subnetwork of $A$ linked to the existing nodes of $B$. Then, the final result of the cascading algorithm
is the fixpoint of the operator
\begin{equation}
G=P_{AB}\circ Q_B\circ P_{AB}\circ Q_A
\end{equation}.
The characterization of the operator $G$ in terms of generating functions 
is the starting point for the current analytical threatments of the $IFM$
model
\cite{BuldyrevNAT2010}; 
it has yet to be investigated wether a 
generalization of such an approach would allow to take into account the 
role of the assortativity. 

In general, even if for simple percolation the method of generating functions 
is still applicable for the case of varying assortativity, the functions to be 
calculated have rarely an analytical closed form. Numerical simulations seem 
therefore to be the main way of investigation to study the effects 
of assortativity in model systems.

For ER and BA networks, IFM exhibit an abrupt transition
\cite{BuldyrevNAT2010} at an {\it a priori\/} unknown value of $x=
x_c$. When a fraction of nodes larger than $1-x_c$ is initially attacked
or fails and hence is removed, the system experiences cascading
disruption and (when the iterative process stops) ends up completely
fragmented, and the relative size of the giant component tends to
vanish.  When a fraction of nodes equal to or less than $1-x_c$ is
attacked or fails, there is always a finite fraction of nodes surviving,
i.e., the giant component relative size does not vanish.

\subsection{Percolation}

In classical percolation \cite{StaufferBook1994}, increasing the
fraction of removed nodes $1-x$ reduces the size (i.e., the number of
nodes) $S$ of the largest cluster.  In the thermodynamic limit
$N\to\infty$ (where $N$ is the number of nodes) the process experiences
a phase transition, i.e., the fraction of nodes $s\equiv S/N$ belonging
to the largest component drops to $s=0$ for $x<x_c$. For $x>x_c$, $s$ is
non-zero.  Depending on the order of the transition, a discontinuous
jump is observed in the order parameter $s$ or in one of its derivatives
with regard to $x$.  Finite size effects round out the behavior of $s$
and make it difficult to distinguish a genuine weak first-order
transition (small jump in $s$) from a second-order transition
\cite{ChallaPT1990}.  Although the sharpness of the transition for the
IFM indicates the possibility that IFM experiences a first-order
percolation transition \cite{BuldyrevNAT2010}, much care must be used in
assessing the transition order for a given network
\cite{daCostaPRL2010}. We want to remark that, while classical percolation
can be described in terms of thermodynamical equilibrium states, this does 
not seem to be tha case for the IFM model. Nevertheless, both percolation and IFM 
share analogous concepts and even techniques (like the use of generating functions 
to produce approximate analytical solutions); therefore, all the standard machinery 
of percolation comes handy in analysing and understanding the IFM.

We simulate the IFM and calculate the size of the largest cluster
(``giant component'') $S$ at varying values of the fraction $x$ of
initially surviving nodes.  As an order parameter for the percolation
transition, we focus on the fraction of nodes $s\equiv S/N$ belonging to
the giant component.  We indicate by $\langle s\rangle$ the value of $s$
averaged both over different layers of the same average assortativity
(the same $J$) and over different IFM simulations.

Next we estimate the percolation threshold $x_c$ corresponding to 
different assortativities (i.e., different $J$ values) using two
methods: 

\begin{itemize}

\item[{\bf I:}] We calculate the point of maximum fluctuation
\begin{equation}
\langle(\delta s)^2 \rangle \equiv\langle s^2 \rangle - \langle s \rangle^2
\label{eq:delta_s}
\end{equation}
of the giant component, which is expected to be large for both first-
and second-order transitions \cite{ChallaPT1990}.

\item[{\bf II:}] In order to compare the estimates of $x_c$ obtained
  by Method I we must also consider the numerical derivative
\begin{equation}
\Delta \langle s \rangle \equiv 
\frac{\langle s(x+\epsilon) \rangle -\langle s(x-\epsilon) \rangle}
{2\epsilon}
\label{eq:Delta_s}
\end{equation}
in the critical region, and we use $\epsilon=10^{-3}$; such a choice 
is dictated from the fact that the derivatives do not show appreciable
numerical changes for smaller values of $\epsilon$.  

In classical percolation $\Delta \langle s \rangle$ 
is equivalent to $\langle(\delta s)^2\rangle$,
\begin{equation}
\Delta \langle s \rangle \approx
\partial_x \langle s \rangle \propto  
\langle(\delta s)^2 \rangle
\label{eq:DsANDds}
\end{equation}
for a second order transition, and it measures the jump $\langle s(x_c^+)\rangle-\langle s(x_c^-)\rangle$ at
$x=x_c$ in the order parameter near a first order transition
\begin{equation}
 \epsilon \Delta \langle s (x_c) \rangle \approx
	\langle s (x_c^+) \rangle - \langle s (x_c^-) \rangle,
\label{eq:DsANDjump}
\end{equation}
where
\begin{equation}
\langle s (x_c^\pm) \rangle \equiv\lim_{\epsilon\to0} \langle s (x_c \pm \epsilon) \rangle
\end{equation}
are the values just before and after the discontinuity.  

\end{itemize}

\noindent Due to the finite size effect, the $x_c$ values found using
these two methods may differ, but as system size increases we expect the
corresponding $x_c$ values from the two methods to converge.

\section{Results}\label{sec:results}

\subsection{Generating the networks}

To generate networks with varying assortativity, we start with a network
with a given degree distribution $P(k)$ and apply MC rewiring for
different values of $J$ according to the sampling probability $\exp
\left[ -J\,H(G) \right]$ \cite{DAgostinoEPL2012}.  Negative values of
$J$ lead to a disassortative network, and positive values to an
assortative network. In other words, when employing a positive $J$,
rewiring connecting nodes of similar degree are accepted more
frequently, whereas when employing a negative $J$, rewiring connecting
nodes of very different degrees are preferred. The absolute value of $J$
behaves like an inverse temperature: the higher its value, the stronger
the selection.  In order to improve the statistics over the configurations, 
we start with uncorrelated initial conditions
(i.e., we restart the procedure from scratch) and generate 100
independent networks for each value of $J$.

We duplicate each configuration to create two topologically identical
monolayers ($A$ and $B$).  Linking each node in layer $A$ to one and
only one node in layer $B$ provides the two layere network systems (2LNS) 
we will employ in our IFM
simulations.  To avoid correlations among the degrees of the two layers,
we first perform a random permutation of the labels of one of the two
layers and then create a connection $A_i \leftrightarrow B_i$ that
represents the mutual dependence of the nodes. For each 2LNS, we perform
100 independent simulations of the IFM model. Thus for each $J$ we
perform $10^4$ simulations starting from 100 different initial networks.

To compare the ER case with the SF case, we generate in both cases 
networks with $N=10,000$ nodes and an average degree $\langle k\rangle=6$.
To generate ER networks (ERnets) with varying assortativity, we employ
17 different $J$ values that produce networks with an average
assortativity that ranges from $r=-0.8$ to $r=0.8$.  To generate
SF networks (SFnets) with varying assortativity, we use the
Barabasi-Albert network \cite{BAreview}, employ eight different $J$
values, and produce networks with an average assortativity that ranges
from $-0.12$ to $0.16$.

\subsection{Breakdown of coupled SF networks}

\begin{figure}
\begin{center}
\includegraphics[width=0.95 \columnwidth]{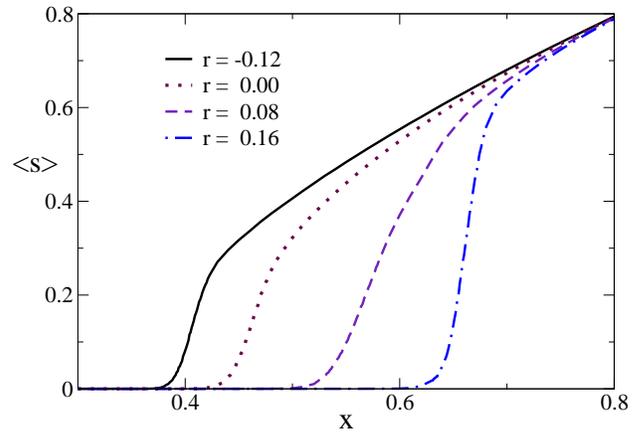}
\end{center}
\caption{(Color online) Fraction $\langle s \rangle$ of sites in the
  giant component as a function of the fraction $x$ of initially
  undamaged nodes for $N=10,000$ scale-free networks.  Curves are
  obtained by averaging 1000 simulations over 100 independent networks for
  each value of $x$.  }
\label{Fig2}
\end{figure}

We simulate the IFM and calculate the fraction of nodes belonging to the
giant component.  Figure \ref{Fig2} shows the behavior of the order
parameter $\langle s\rangle$ as a function of the fraction of survived
nodes for the SFnets.  Note that the size of the giant component
increases significantly in a limited region that depends on the
assortativity; in a system of inite size, this is an indication 
for a percolation phase
transition.  Two different regions of stability can be identified that
correspond to the two different phases, (i) a percolative 
phase in which
the giant component includes a number of nodes proportional to $N$
($S\sim N$, i.e., a non-vanishing $s$) and (ii) a broken phase in
which the largest component is negligible ($S\sim o(N)$, i.e., $s\sim
0$).  The amount of damage needed to destroy the giant component
decreases with assortativity, and we find the sharpness of the
transition to decrease at fixed system size. 

\hl{Such an effect can be understood by observing that the 
breakdown process consists in repeated applications of a 
percolation algorithm on single networks. In the case of a 
single network} \cite{NewmanPRL2002} \hl{increasing the assortativity
reduces the extension of the largest component. In other
words, at each iteration, the fraction of removed nodes (complementary 
to the giant component) increases; thus, the iterations over the two
networks amplify the effect of the assortativity easing the breakdown 
of the coupled system.}

\begin{figure}
\begin{center}
\includegraphics[width=0.95\columnwidth]{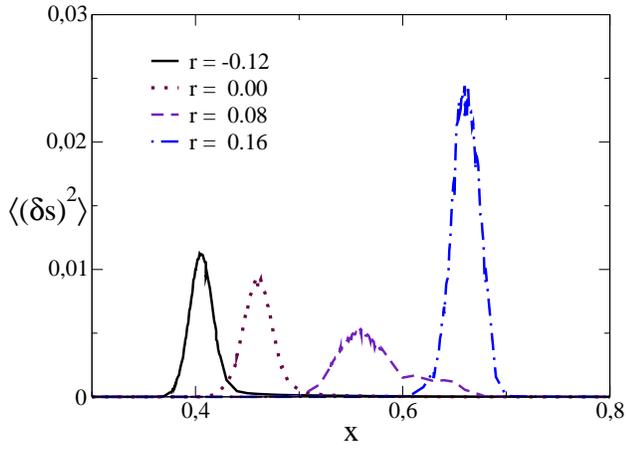}
\end{center}
\caption{(Color online) Fluctuations $\langle (\delta s)^2 \rangle$ of
  the order parameter $s$ as a function of the fraction $x$ of the
  initially undamaged nodes for $N=10,000$ scale-free networks.  The
  position of the peak for $\langle (\delta s)^2 \rangle$ can be used to
  estimate the critical fraction of non-damaged sites $x_c$.  Curves are
  obtained by averaging 1,000 simulations over 100 independent networks
  for each value of $x$.  }
\label{Fig3}
\end{figure}

To estimate the percolation threshold $x_c$ that corresponds to
different $J$ values (i.e., to the different assortativities $r$), we
calculate the point of maximum fluctuation for the size of the giant component.
Figure~\ref{Fig3} shows the fluctuations of the largest component
$\langle(\delta s)^2 \rangle$ in a SFnet as a function of $x$ for sample
values of the SFnet. To attain a better estimate $x_c$, we perform more
simulations in the region where the maximum 
of $\langle(\delta s)^2 \rangle$ is attained.

\begin{figure}
\begin{center}
\includegraphics[width=\columnwidth, angle =0]{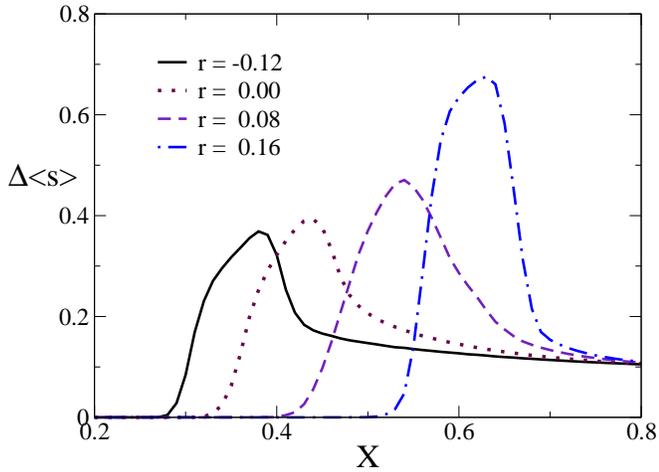}
\end{center}
\caption{(Color online) Normalized numerical increment $\Delta \langle
  s \rangle$ of the order parameter $s$ as a function of the fraction
  $x$ of initially undamaged nodes for $N=10,000$ scale-free
  networks. In conventional percolation $\Delta\langle
  s\rangle\sim\langle\delta s\rangle$ can be used to estimate the the
  critical fraction of damaged sites $x_c$.  Curves are obtained by
  averaging 1,000 simulations over 100 independent networks for each
  value of $x$.  }
\label{Fig4}
\end{figure}

Figure \ref{Fig4} shows that the numerical derivative $\Delta\langle
s\rangle$ also shows a peak in the critical region.  Note that one may
also estimate the critical threshold as the inflection point of the
largest component profiles, i.e., from the peak of the numerical
derivative $\Delta \langle s \rangle$.  Nevertheless, no significant
difference is observed within the accuracy of our simulations, i.e., the
inflection points of $\langle s \rangle$ coincide, within the error
bars, with a maximum of $\langle (\delta s)^2 \rangle$.

\begin{figure}
\begin{center}
\includegraphics[width=0.95\columnwidth]{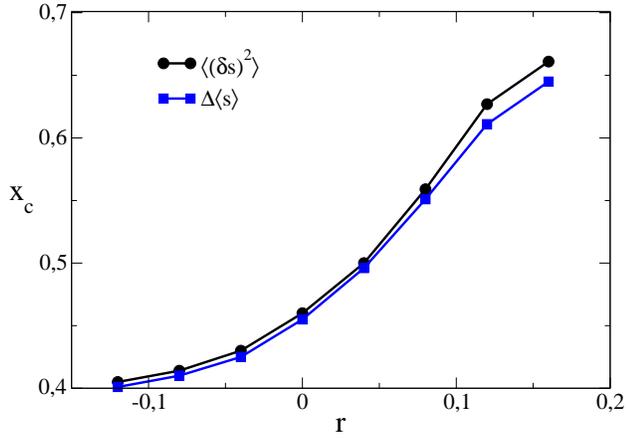}
\end{center}
\caption{Estimated values of the percolation threshold $x_c$ as a
  function of the assortativity coefficient $r$ for $N=10,000$
  scale-free networks.  The values of $x_c$ are estimated as the maxima
  of $\langle (\delta s)^2 \rangle$, as well as the peaks of the 
  $\Delta \langle s \rangle$ profiles. Note that the two estimates are very
  close for disassortative nets but differ a bit more for assortative
  nets.  }
\label{Fig5}
\end{figure}

Using the two methods above (via the peak of $\langle (\delta s)^2
\rangle$ or via the peak of $\Delta \langle s \rangle$) we can obtain
the dependence of the percolation threshold $x_c$ on the assortativity
$r$.  Figure~\ref{Fig5} shows data for the SFnets and provides evidence
that the percolation threshold is an increasing function of the
assortative coefficient $r$; 
therefore, robustness decreases with increasing assortativity.

\begin{figure}
\begin{center}
   \includegraphics[width=0.95\columnwidth]{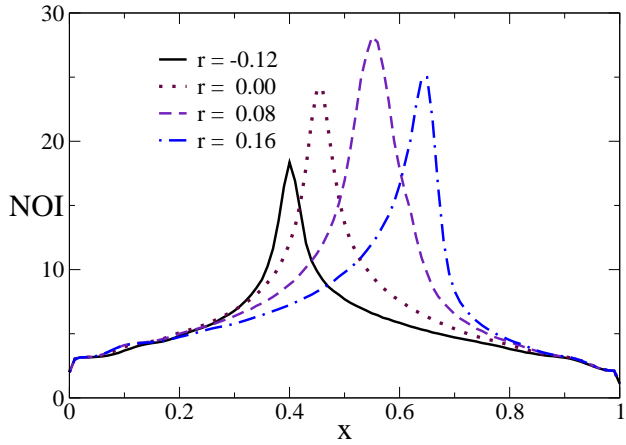}
\end{center}
\caption{(Color online) Number of iterations (NOI) for the IFM algorithm
  to converge as a function of the initially undamaged node
  fraction. Peak positions for different assortative coefficients are
  close to $x_c$ as estimated from $\langle (\delta s)^2 \rangle$.}
\label{Fig6}
\end{figure}

As a general result, it has been observed that a phase transition in a
numerical model often coincides with a peak in the number of operations
required to calculate the significant quantities (order parameters and
potentials). In our case the number of iterations (NOI) needed for the IFM
algorithm to converge represents a natural measure for the
computing operations.  Consistent with the general principle,
Fig.~\ref{Fig6} shows that the NOI for the IFM algorithm exhibits a peak
close to the critical threshold.  As a possible interpretation, we note
 that the NOI represents the sum of a set of stochastic
variables (one for each iteration) that are null when the removal of
nodes does not fragment the giant component and are unitary
elsewhere. Therefore the NOI measures the stability of the largest
component upon further node removal.

\begin{figure}
\begin{center}
   \includegraphics[width=0.95\columnwidth]{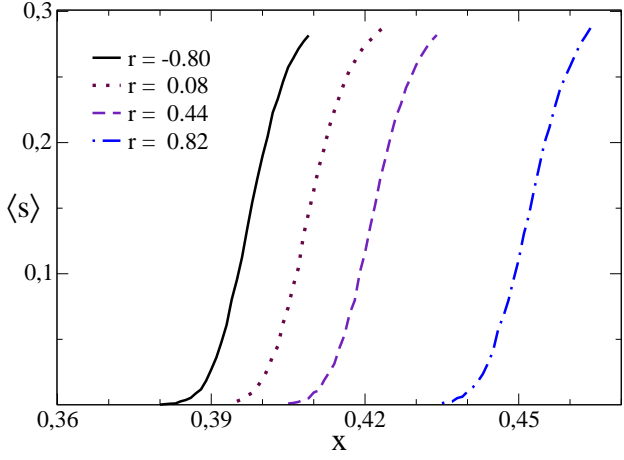}
\end{center}
\caption{(Color online) Fraction $\langle s \rangle$ of sites in the
  largest component as a function of the fraction $x$ of initially
  undamaged nodes for $N=10,000$ Erd\H{o}s-R\'{e}nyi networks.  Curves
  are obtained by averaging for each $x$ value $10^4$ simulations over
  $100$ independent networks.  }
\label{Fig7}
\end{figure}

\begin{figure}
\begin{center}
   \includegraphics[width=0.95\columnwidth]{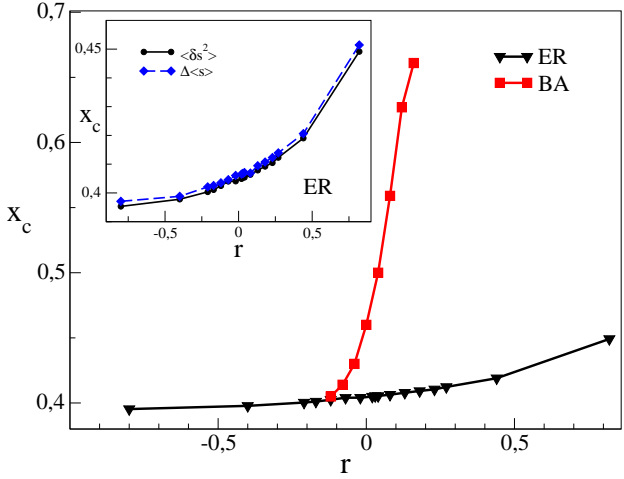}
\end{center}
\caption{(Color online) Comparison of the percolation thresholds $x_c$
  for both the Erd\H{o}s-R\'{e}nyi and scale-free interacting networks.
  Scale-free networks exhibit a significant variation upon a small
  increase of the assortativity $r$, but Erd\H{o}s-R\'{e}nyi networks
  exhibit only a small variation over the whole possible range of
  assortativities.  Inset: the estimates of $x_c$ via the normalized
  increment $\Delta \langle s \rangle$ are very close to the estimates
  of $x_c$ via the peaks of $\langle (\delta s)^2 \rangle$ also for
  Erd\H{o}s-R\'{e}nyi networks.}
\label{Fig8}
\end{figure}

As mentioned above, we generate ER networks with the same average
degree.  Thus configurations with a given size $N$ and a given
assortativity $r$ are distinguishable only by their degree
distributions.  So if we plot the properties of networks of the same
size versus $r$, we can pinpoint and compare the difference between the
behavior of ER and SF networks.  In Fig.~\ref{Fig7} we show that the
phase transition requires an increasing number of damaged sites with
increasing assortativity for ERnets as well.  Unlike SFnets, the effect
on the critical threshold in ERnets is much more limited. In
Fig.~\ref{Fig8} we compare the estimated thresholds in the two cases.
It is clear that in a SFnet the critical threshold $x_c$ varies
dramatically, but that in a ERnet it is almost flat.
\hl{This effect of enhanced response to assortativity in the SFnets 
with respect to the ERnets is consistent to what is observed in 
single layer networks} \cite{DAgostinoEPL2012}.

\subsection{Order of the phase transition}        

It is difficult to determine the order of a phase transition from
simulations. To see whether this kind of phase transition is first-order
or second-order, we analyze the fluctuations of the size of the giant component.

\begin{figure}
\begin{center}
   \includegraphics[width=0.95\columnwidth]{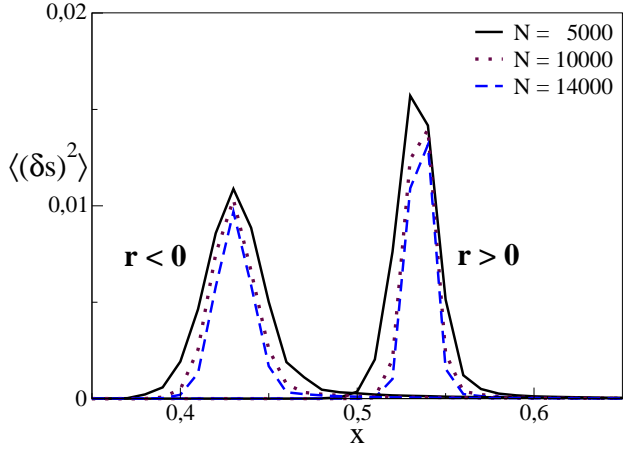}
\end{center}
\caption{(Color online) Fluctuations $\langle (\delta s)^2 \rangle$ of
  the order parameter $s$ as a function of the fraction $x$ of initially
  undamaged nodes for scale-free networks of different sizes.  The peaks
  on the left correspond to disassortative networks ($J=-10$, i.e., $r$
  from $-0.086$ to 0.051) while the peaks on the right correspond to
  assortative networks ($J=10$, i.e., $r=$ from 0.181 to 0.163).  The
  different curves correspond to sizes $N=5,000$ (circles), $N=10,000$
  (squares), and $N=14,000$. For the sizes analysed, there is no
  significant evidence for the growth and narrowing of the peaks that
  would be expected in a second-order transition.}
\label{Fig9}
\end{figure}

In fact, for a second order transitions the divergence of the
fluctuations $\langle (\delta s)^2 \rangle$ at $x_c$ would be signaled at 
finite system sizes by an increase in the peak of $\langle (\delta s)^2
\rangle$ and by the narrowing of the width of the peak.  
In Fig.~\ref{Fig9} we show that, although there is a slight narrowing of
the peaks with system size, there is no sign of a second
order-divergence.  Figure~\ref{Fig7} shows the fluctuation profiles for
SFnets for different sizes ($N=5,000$, 10,000, and 14,000). We find
analogous results for ERnets.

To further check whether this kind of phase transition is first-order
or second-order, we analyze the size of the second largest cluster $S_2$
and its counterpart $s_2 \equiv S_2/N$.  In a second order transition, 
the presence of a sharp peak in $s_2$ is coupled to a sharp increase 
of $s$ near $x=x_c$.  In fact, at criticality, the size
of the second percolating cluster has the same scaling as the giant
component \cite{NaemJSP1998} also for systems above the critical dimension
\cite{daSilvaPRE2002}.  On the other hand, first order transitions are
characterized by finite size clusters and by $S_2/N \to 0$.

The second largest cluster is found by following the algorithm
\ref{alg:IFM} but starting from the second largest cluster (instead of
the largest) at the first iteration.  When we vary $x$ in our
simulations we do not observe a peak in $s_2$.  The size of the second
largest cluster is flat and always of order $\sim 1/N$.

Therefore, all these arguments support the presence of a first-order transition 
as predicted by mean-field ER network-of-networks models \cite{GaoPRL2011,GaoNAT2012}.

Let us finally comment on the role of the hysteresis in signalling the order of the 
transition. In a first order transition, local minima of the free energy develop before 
the transition point and become the favoured one at the transition point; on the same 
footings the local free energy minimum corresponding to the equilibrium state before the transition persists as a metastable state for some range of the parameters. 
The switch of the favoured minima
at the transition point is signalled by a jump in any macroscopic quantities that 
discriminates among such minima; nevertheless, if minima are deep enough, the system can 
persist in the metastable state for a finite time before jumping to equilibrium.
Such behaviour results in hysteresis curves and is therefore linked to the equilibrium
description of the system in terms of free energy.
In the IFM there is not a free energy description of the system, but 
a characterization of the final state as fix-points of an operator: therefore the 
applications of statistical mechanics observables is just a guidance in studying 
the system; on the other hand, distinguishing the cases where the transition is 
abrupt (first order) is of great interest and importance for real systems.

\section{Conclusion}\label{sec:conclusions}

We have examined the influence of assortativity on the robustness of
interdependent systems consisting of two interacting networks.  Both
scale-free (SF) and Erd\H{o}s-R\'{e}nyi (ER) network models have been
taken into account. The simulation of cascading faults caused by random
attack or failure in the interdependent pair of networks provides
evidence for a first order percolation transition.

The percolation threshold decreases with increasing assortativity and
therefore assortative networks are more fragile in both the ER and SF
cases but, generally speaking, SF networks are less robust than ER
interdependent pairs. Even a low assortativity can make a SF network
100\% more fragile than its corresponding ER version.

\acknowledgments

AS acknowledges support from FET Open Project ``FOC'' nr. 255987 and from DTRA.  
GD acknowledges support by the European project ``MOTIA''
JLS-2009-CIPS-AG-C1-016.  DZ thanks DTRA, ONR, and NSF (grant
CMMI-1125290) for support.

\bibliographystyle{apsrev}
\bibliography{twolayers}

\newpage 

\end{document}